\documentclass[aps,pra,showpacs,amsmath,amssymb,superscriptaddress,reprint,10pt]{revtex4-1}

\usepackage{graphicx}
\usepackage{dcolumn}
\usepackage{bm}
\usepackage[latin1]{inputenc}
\usepackage{epstopdf}
\usepackage{amsmath}
\usepackage{upgreek}  %<-- Use \upmu to avoid capital italicized Greek letters
\usepackage{amssymb}
\usepackage{esint}
\usepackage{array}
\usepackage{xcolor}
\usepackage{latexsym}
\usepackage{mathrsfs}
\usepackage[sans]{dsfont}
\usepackage[colorlinks=true,breaklinks=true,allcolors=blue]{hyperref}

\begin{document}

\preprint{}

\title{A compact experimental machine for studying tunable Bose-Bose superfluid mixtures}

\author{P.C.M. Castilho}
\email{patricia.castilho@usp.br}
\altaffiliation{Current address: Laboratoire Kastler Brossel, Coll\`{e}ge de France, CNRS, 11 Place Marcelin Berthelot, 75005 Paris, France}
\affiliation{Instituto de F\'{\i}sica de S\~{a}o Carlos, Universidade de S\~{a}o Paulo, C.P. 369, 13560-970 S\~{a}o Carlos, SP, Brazil}
\author{E. Pedrozo-Pe\~{n}afiel}
\altaffiliation{Current address: Department of Physics, MIT-Harvard Center for Ultracold Atoms and Research Laboratory of Electronics, Massachusetts Institute of Technology, Cambridge, Massachusetts 02139, USA}
\affiliation{Instituto de F\'{\i}sica de S\~{a}o Carlos, Universidade de S\~{a}o Paulo, C.P. 369, 13560-970 S\~{a}o Carlos, SP, Brazil}
\author{E.M. Gutierrez}
\affiliation{Instituto de F\'{\i}sica de S\~{a}o Carlos, Universidade de S\~{a}o Paulo, C.P. 369, 13560-970 S\~{a}o Carlos, SP, Brazil}
\author{P. L. Mazo}
\affiliation{Instituto de F\'{\i}sica de S\~{a}o Carlos, Universidade de S\~{a}o Paulo, C.P. 369, 13560-970 S\~{a}o Carlos, SP, Brazil}
\author{G. Roati}
\affiliation{INO-CNR and LENS, University of Florence, via N. Carrara 1, 50019 Sesto Fiorentino, Italy}
\author{K.M. Farias}
\affiliation{Instituto de F\'{\i}sica de S\~{a}o Carlos, Universidade de S\~{a}o Paulo, C.P. 369, 13560-970 S\~{a}o Carlos, SP, Brazil}
\author{V.S. Bagnato}
\affiliation{Instituto de F\'{\i}sica de S\~{a}o Carlos, Universidade de S\~{a}o Paulo, C.P. 369, 13560-970 S\~{a}o Carlos, SP, Brazil}

\begin{abstract}
We present a compact and versatile experimental system for producing Bose-Bose superfluid mixtures composed of sodium and potassium atoms. The compact design combines the necessary ultra-high vacuum enviroment for ultracold atom experiments with efficient atomic fluxes by using two-dimensional magneto-optical traps as independent source of atoms. We demonstrate the performance of this new machine by producing a Bose-Einstein condensate of $^{23}$Na with $\sim 1 \times 10^{6}~$atoms. The tunability of Na-K bosonic mixtures is particularly interesting for studies regarding the nucleation of vortices and quantum turbulence. In this direction, the large optical access of the science chamber along the vertical direction provides the conditions to implement high resolution optical setups for imaging and rotating the condensate with a stirring beam. We show the nucleation of a vortex lattice with up to 14 vortices in the $^{23}$Na BEC, attesting the efficiency of the experimental apparatus in studying the dynamics of vortices.
\end{abstract}

%\pacs{42.25.Fx, 32.80.Pj}
\maketitle

\section{Introduction}\label{Introduction}

Quantum degenerate mixtures created with either different Zeeman sublevels~\cite{Myatt1997TwoBECs87Rb,Hall1998DynamicsTwoBECs87Rb,Stenger1998SpinorBECNa} or with different atomic species~\cite{Modugno200287Rb41K,Hadzibabic20026Li23Na,Modugno200287Rb40K,Silber20056Li87Rb, Ospelkaus2006Tunning40K87Rb,Papp2008TunableMiscibility87-85Rb,Taglieber2008FermiFermi, McCarron2011BEC87Rb133Cs,Pasquiou2013SrRb,Farrier-Barbut2014BoseFermiSuperfluids,Wacker2015Tunable87Rb39K, Park2015UltracoldDipolarMol,Wang2016Tunable23Na87Rb,Schulze2017Tunable23Na39K} exihibit a rich physics that is not accessible in a single component quantum gas. The existence of different miscibility regimes~\cite{Stenger1998SpinorBECNa,Ospelkaus2006Tunning40K87Rb,Papp2008TunableMiscibility87-85Rb, McCarron2011BEC87Rb133Cs,Wacker2015Tunable87Rb39K,Wang2016Tunable23Na87Rb,Schulze2017Tunable23Na39K, Lee2016PhaseSeparationTwoBECs} directly affects the statistical and dynamical properties of the system. The nucleation of vortices~\cite{Kasamatsu2003RotTwoBECs,Mason2011RotTwoBECs,Kuopanportti2012ExoticVortexLattices} and the superfluid current stability~\cite{Smyrnakis2009SupercurrentTwoBECs,Anoshkin2013CurrentsBosonicMix,Beattie2013CurrentSpinorBEC} are strongly modified by the interplay between the inter- and intraspecies interactions. In optical lattices, the phase diagram of two-component systems is far more complex than the simple extension of the standard superfluid to Mott insulator transition~\cite{Altman2003TwoBosonsOptLattice,Isacsson2005Superfluid-Insulator}. Aditionally, the possibility of producing heteronuclear ground-state molecules with large dipole moment~\cite{Park2015UltracoldDipolarMol,Carr2009UltracoldMolecules,Molony2014Molecules87Rb133Cs} could enable the study of even stronger interacting dipolar gases than the ones recently obtained with lanthanide atoms~\cite{Aikawa2012BECEr,Kadau2016RosensweigInstability}. Large imbalanced mixtures can shed light on the physics of impurities coupled to a bosonic bath~\cite{Bruderer2008SelfTrappingImpurities,Santalore2011PolaronsBEC}, of fundamental interest in condensed matter physics. And beyond mean-field effects could be explored with the recent observation of quantum droplets in Bose mixtures composed of two hyperfine states of $^{39}$K~\cite{Cabrera2017Droplets39K,Semeghini2017Droplets39K}, as it was originally proposed in~\cite{Petrov2015Droplets}.

The specific case of Na-K mixtures is of particular interest due to its large flexibility. Both, Bose-Bose and Bose-Fermi mixtures could be easly produced by changing the potassium isotope and the different miscibility regimes explored with the tunning of the interspecies interactions via the Feshbach resonances for each combination~\cite{Viel2016NaKFeshPredictions}. Due to the existence of chemically stable ground-state molecules with large eletric dipole moment ($\sim 2.72~$Debye), the Bose-Fermi mixture has been vastly explored during the last few years~\cite{Park2015UltracoldDipolarMol}.

In this work, we present the realization of a compact experimental machine to produce Bose-Bose superfluid mixtures composed of $^{23}$Na and $^{39}$K or $^{41}$K atoms with fast repetion rate. While the combination with $^{39}$K, recently produced in~\cite{Schulze2017Tunable23Na39K}, could become a good platform for the investigation of bosonic impurities, the combination $^{23}$Na-$^{41}$K represents a fully-tunable Bose-Bose superfluid which remains still now unexplored. Thanks to its tunability, the nucleation of coupled vortex lattices~\cite{Mason2011RotTwoBECs,Kuopanportti2012ExoticVortexLattices} and binary quantum turbulence~\cite{Takeuchi2010BinaryQT,Tsatsos2016ReviewQT} configures one promissing path for such mixture. Moreover, the large difference between their atomic transitions (with $\lambda_{\rm Na}=589~$nm and $\lambda_{\rm K}=767~$nm) makes it possible to selectively nucleate vortices in one species even without a \textit{magic wavelength}. This could be done with the use of an optical potential (i.e. stirring beam~\cite{Madison2000RotatingBEC,AboShaeer2001VortexLattice,Raman2001StirringBEC} and moving barrier) at $532~$nm, which affets strongly sodium than potassium. The unperturbed species could be used to map the first species evolution revealling the tangling between vortex filaments. Similar experiments had been developed in superfluid helium~\cite{Bewley2006MappingVorticesHe,Guo2010CounterflowHe}, but no analogous have been performed in quantum gases so far. In the following, we focus on describing the experimental apparatus and the experimental sequence for producing a vortex lattice in a Bose-Einstein condensate of sodium atoms after rotation by the stirring beam.

This paper is structured as follows. Section~\ref{sec:NaKSetup} describes the design of the new experimental system including the details of the vacuum system in Sec.~\ref{subsec:VacuumSystem}, of the laser systems for sodium and potassium atoms in Sec.~\ref{subsec:LaserSystems} and of the conservatives traps designed for simultaneously trap the two atomic species in Sec.~\ref{subsec:Traps}. The magneto-optical traps (MOTs) for each species are described in Sections~\ref{subsec:Na2De3DMOTs} and~\ref{subsec:K2De3DMOTs} followed by the characterization of the combined two-species MOT, in~\ref{subsec:TwoMOTs}. Later, in Section~\ref{sec:NaBEC}, we focus on describing the experimental sequence for producing the $^{23}$Na BEC. The intermediate steps used to improve the transference of the atoms from the 3D-MOT to the optically plugged Quadrupole trap are described in Sec.~\ref{subsec:MOTtoPlugTrap}. Next, the Plug trap and the RF-forced evaporation of the sodium atoms are characterized in Sec.~\ref{subsec:NaPlugTrap}. In Sec.~\ref{subsec:NaBECODT}, we characterize the $^{23}$Na BEC obtained in the crossed optical dipole trap. Finally, a high resolution imaging and stirring setup is described in Sec.~\ref{subsec:ImgSetup} and, in Sec~\ref{subsec:VortexLattice}, we presente the nucleation of a vortex lattice in the $^{23}$Na BEC.

\section{Experimental setup for producing the Na-K Bose-Bose superfluid mixtures}\label{sec:NaKSetup}

In this section, we describe the design of each common experimental part in producing the Na-K Bose-Bose superfluid mixtures.

\subsection{A compact two-species vacuum system}\label{subsec:VacuumSystem}

Cold atom experiments require the combination of an ultra-high vacuum environment (with $P\sim 10^{-11}-10^{-12}~$Torr) and efficient atomic sources. In the past years, compact sources of atoms have gained great attention due to the high atomic fluxes obtained from two-dimensional magneto-optical traps (2D-MOTs). These traps, initially implemented for rubidium and potassium isotopes in a vapor cell~\cite{Dieckmann19982DMOTRb, Catani20062DMOTK}, were later also able to produce high atomic fluxes of lithium and sodium atoms by adding an adapted ``Zeeman slower''~\cite{Phillips1982ZeemanSlower} which pre-cool the atoms coming from an oven~\cite{Tiecke20092DMOTLi, Lamporesi20132DMOTNa}. For the latter atomic species, a direct comparison between the loading of a standard three-dimensional MOT (3D-MOT) from pre-cooled atoms coming from a real Zeeman slower and from the \textit{modified} 2D-MOT has shown the equivalence of the two atomic fluxes~\cite{Pedrozo2016Comparison}.

\begin{figure}\centering
\includegraphics[width=0.48\textwidth]{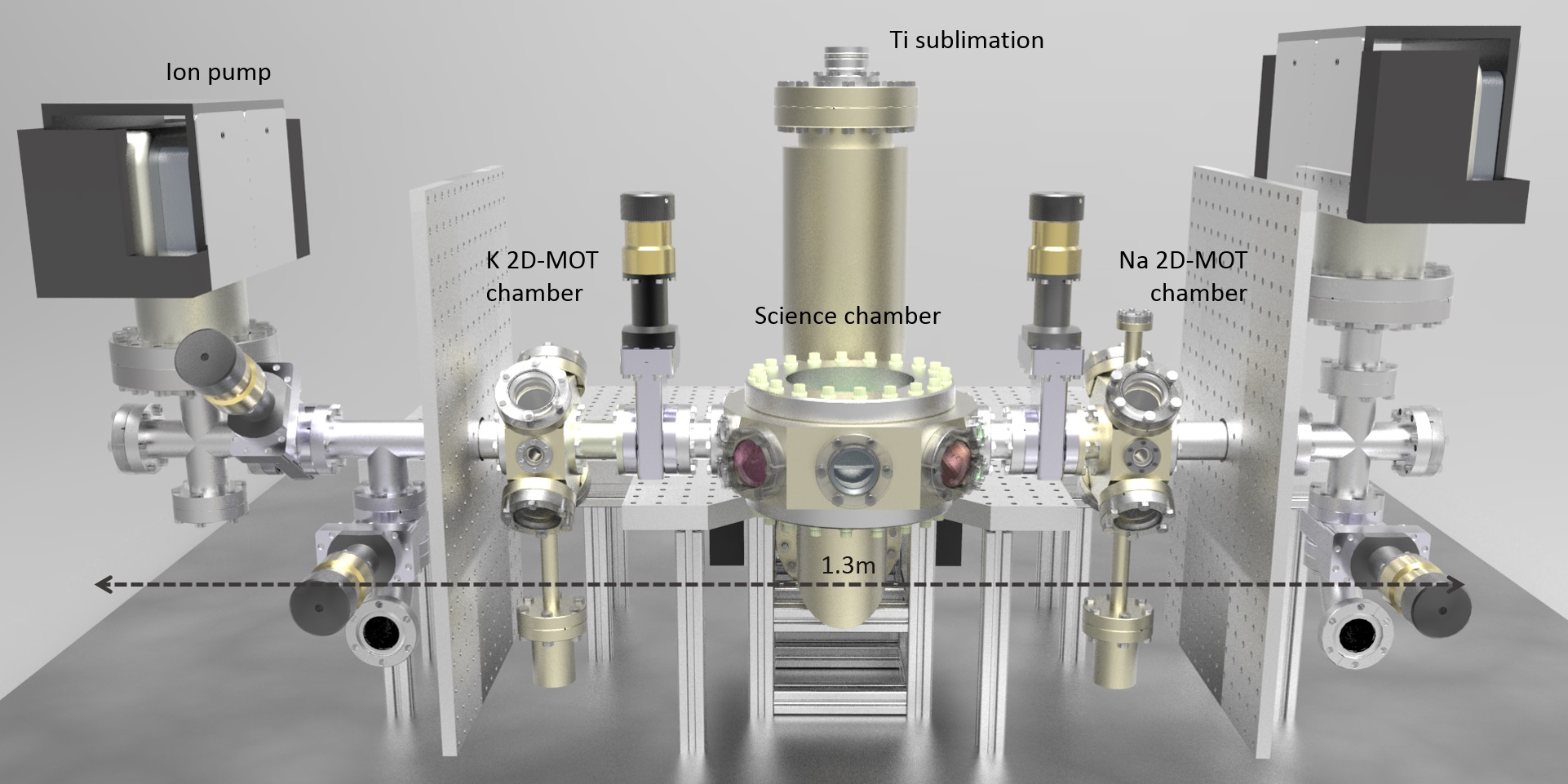}
\caption{
Front view of the compact vacuum system designed for the production of the Na-K Bose-Bose superfluid mixtures.
}
\label{fig:NaKVacuumSystem}
\end{figure}

The compact vacuum system designed for the production of the Na-K Bose-Bose superfluid mixtures uses 2D-MOTs performed in independent vacuum chambers as source of atoms for each atomic species. A sketch of the experimental setup is displayed in figure~\ref{fig:NaKVacuumSystem}. The 2D-MOT vacuum chambers (on the sides of the figure) are connected to the Science chamber (SC, in the center) from opposite sides by narrow channels which ensure a differential pumping of $10^{-3}$ and $10^{-4}$ with respect to the 2D-MOT chamber of sodium and potassium, respectively. Three ion pumps (Varian Vaclon Plus $75~$L/s) are installed in the vacuum system pumping each of its three regions in order to keep the desired ultra-high vacuum environment in the SC. The complete vacuum system is only $1.3~$m long which is comparable with the usual sizes of Zeeman slowers used to cool sodium~\cite{Pedrozo2016Comparison}.

The design of the 2D-MOT chambers was inspired by~\cite{Lamporesi20132DMOTNa}: they consist of a one piece metallic chamber made of stainless steel 316L in a X-shape allowing for optical access. A $100~$mm long tube with internal diameter of $17~$mm connects the bottom of the 2D-MOT chambers to an oven containing $5~$g ampole of metallic sodium or potassium. The ovens are heated to usual operating temperatures of $240^{\circ}$C for sodium and $30^{\circ}$C for potassium resulting in a background pressure of $5\times10^{-9}~$Torr and $1\times 10^{-9}~$Torr, respectively. The 2D-MOT for potassium (K 2D-MOT) is loaded from the background vapor filling the whole chamber which is kept at $\sim 42^{\circ}$C. Differently from the K 2D-MOT, a \textit{modified} 2D-MOT with the addition of an adapted Zeeman slower is used to cool the sodium atoms such that the Zeeman beam enters through the upper viewport which is kept at $155^{\circ}$C avoiding deposit of sodium.

The Science chamber also consists of a one piece metallic chamber made of stainless steel 316L. It was specially designed in an almost cylindrical shape with height of $77~$mm and maximum radius of $120~$mm in order to maximize the optical access. Therefore, besides the two 2D-MOT chambers, six CF35 viewports were placed at the $xy$-plane. Eight CF16 viewports were also placed on the same side but with an angle along the vertical direction that orients them to the center of the chamber. Finally, two custom designed re-entrant viewports from Torr Scientific with $89~$mm of clear aperture were placed along the vertical direction with an internal separation of $31~$mm. The resulting large numerical aperture enables the design of a high resolution imaging system along gravity. Besides the ion pump, a titanium sublimation pump (Agilent - Titanium Sublimation Cartridge 916-0050) is connected in the SC region (see cylinder in the center of figure~\ref{fig:NaKVacuumSystem}) enabling the achievement of an ultra-high vacuum environment.

\subsection{Laser systems}\label{subsec:LaserSystems}

In this part, we briefly describe the laser systems designed for sodium and potassium atoms.

\subsubsection{Sodium laser system}

\begin{figure}\centering
\includegraphics[width=0.49\textwidth]{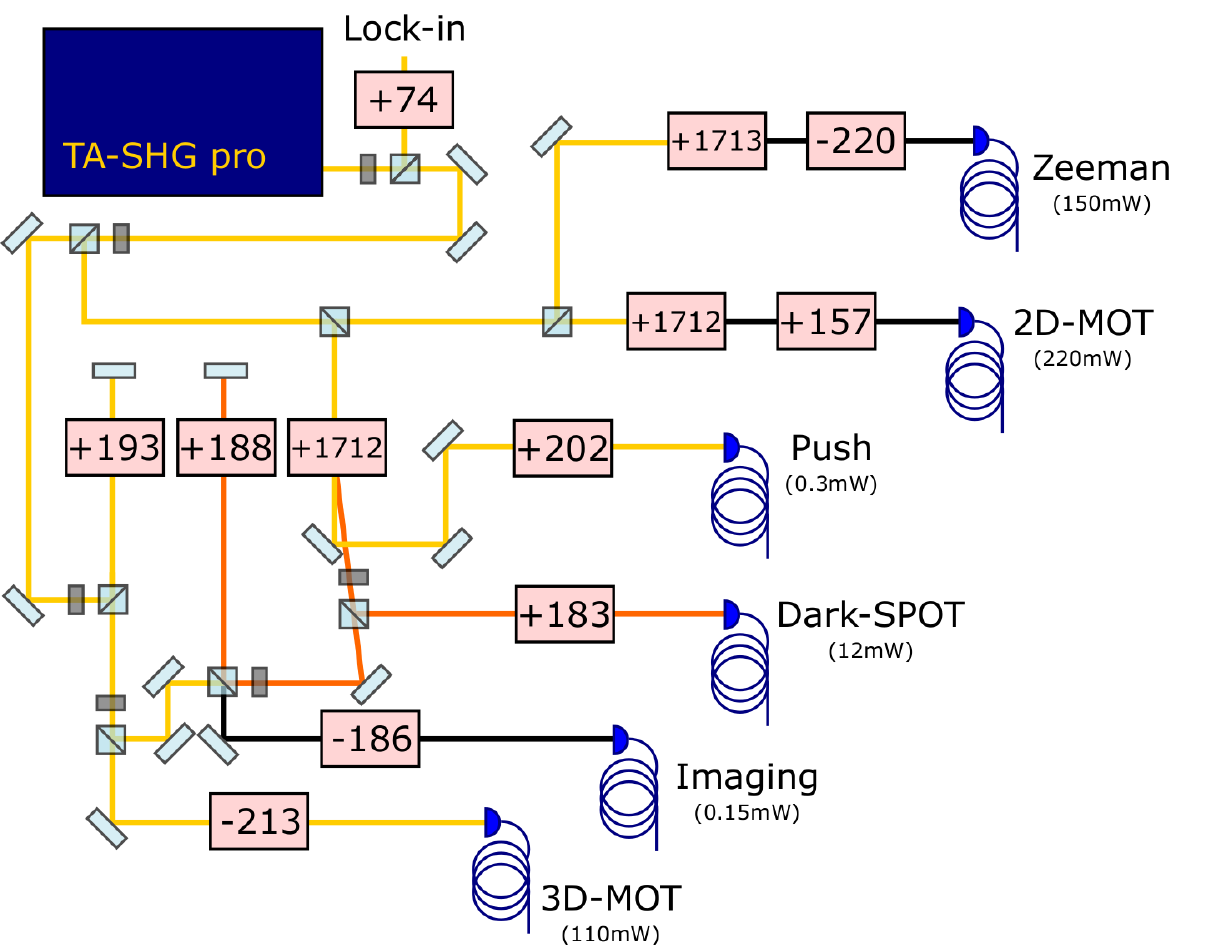}
\caption{
Sketch of the optical setup designed to produce the laser light used to cool, manipulate and image the sodium atoms. The yellow lines represent the laser beams with frequency near the \textit{cooling} transition, the orange lines represent the laser beams with frequency near the \textit{repumper} transition and the black lines represent laser beams with both frequencies. The green line represents the green light at $532~$nm used to pump the dye laser in order to produce the yellow light at $589~$nm. The frequencies used in each AOM/EOM are given in MHz.
}
\label{fig:NaLaserSystem}
\end{figure}

The laser light necessary for cooling, imaging and manipulating the sodium atoms is obtained from a doubly-cavity laser (model DL-RFA-TA-SHG, from Toptica) stabilized to the $3^{2}{\rm S}_{1/2}\vert F=2\rangle\rightarrow 3^{2}{\rm P}_{3/2}$ transition. In figure~\ref{fig:NaLaserSystem}, we present the layout of the optical setup for the sodium laser system.

\begin{figure*}\centering
\includegraphics[width=0.80\textwidth]{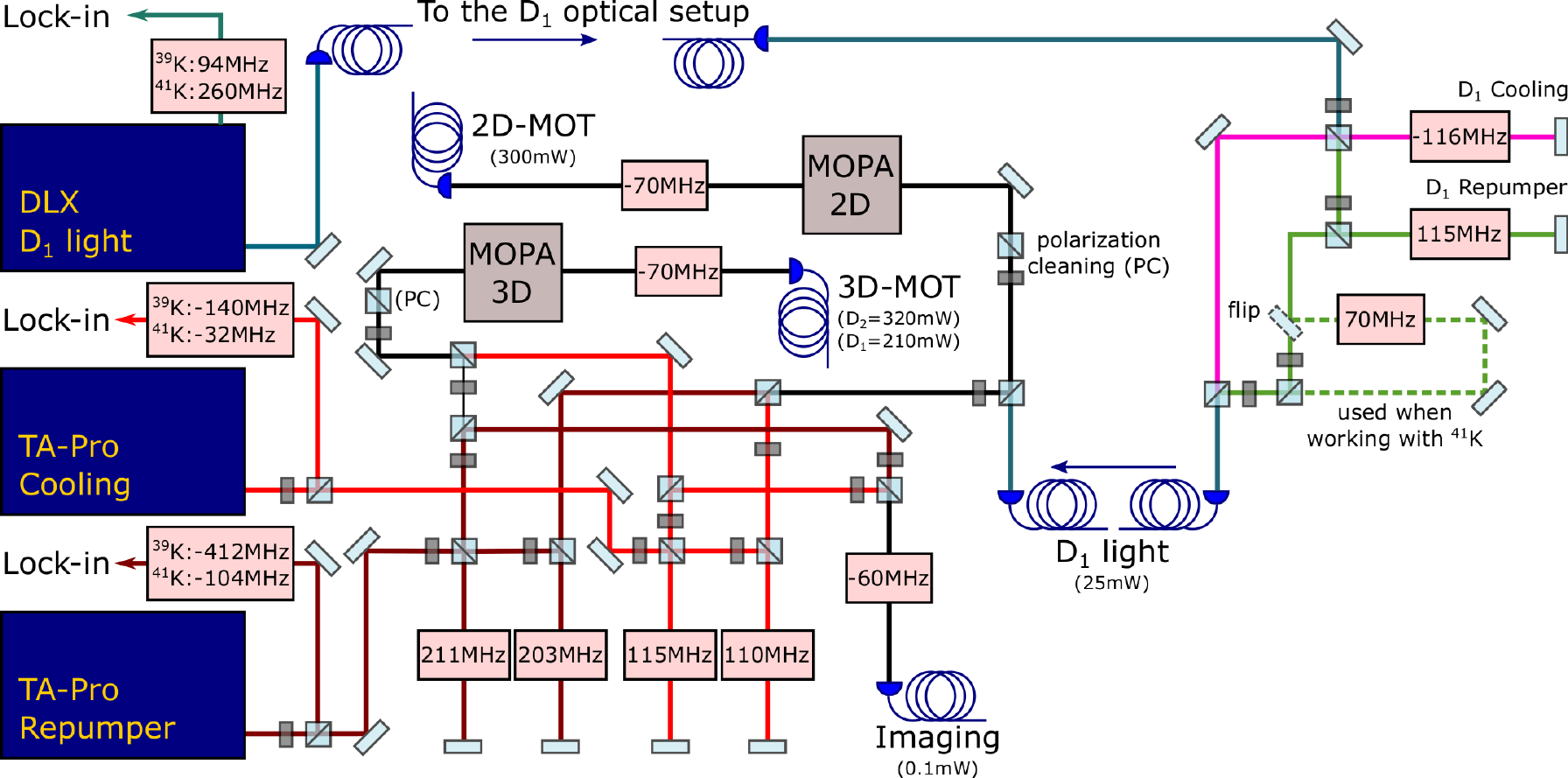}
\caption{
Sketch of the optical setup designed to produce the laser light used to cool, manipulate and image the potassium atoms. The light red lines represent the laser beams with frequency near the $D_{2}$ \textit{cooling} transition, dark red lines represent laser beam with frequency near the $D_{2}$ \textit{repumper} transition, the pink lines represent the laser beam with frequency near the $D_{1}$ \textit{cooling} transition, the light green represent laser beams with frequency near the $D_{1}$ \textit{repumper} transition and the black lines represent laser beams with more than one frequency. The frequencies used in each AOM are given in MHz.
}
\label{fig:KLaserSystem}
\end{figure*}

Electro and acousto-optic modulators (EOMs and AOMs) are used to tune the different laser frequencies and, in the case of the second, also as fast switchers. The laser light is initially divided into four laser beams. The first generates the \textit{cooling} light for the 3D-MOT and the \textit{imaging} light resonant with the $\vert F=2\rangle\rightarrow\vert F'=3\rangle$ transition. A double-pass AOM is used to perform the fine frequency adjustment needed during the experimental sequence with minor misalignment of the laser beams. The second beam, used to generate the \textit{repumper} light of the 3D-MOT and the \textit{pump} light resonant with the $\vert F=1\rangle\rightarrow\vert F'=2\rangle$ transition, is obtained after passing through a shifter acting at $\sim 1.712~$GHz. High frequency optical devices usually present very low efficiency ($\approx 30\%$) and a lot of power remains on the non-diffracted beam. We circumvent this problem by using its 0$^{\rm th}$-order to produce the light needed for the \textit{push} beam used to transfer the pre-cooled atoms at the 2D-MOT to the Science chamber. The third beam is used to generate the \textit{cooling} and \textit{repumper} lights necessary for the 2D-MOT. This is done by passing the light through an EOM with side-bands at $\sim 1.712~$GHz. Finally, the fourth beam is responsible for performing the adapted Zeeman slower in the Na 2D-MOT. The \textit{repumper} light is obtained by the use of an EOM with side-bands at $1.713~$GHz and an AOM at $220~$MHz is used to bring both lights far red detuned from the atomic transition.

Polarization maintaining fibers deliver all the manipulated lights on the experimental apparatus decoupling the laser sources from the optical table containing the vacuum system. On the experiment table, all laser beams are adjusted to have a beam diameter of $20~$mm, besides the push beam, which has a beam diameter of $3~$mm. The usual output laser powers for each fiber are specified in figure.~\ref{fig:NaLaserSystem}. 

\subsubsection{A versatile laser system for bosonic potassium isotopes}

The bosonic potassium isotopes ($^{39}$K and $^{41}$K) have transition frequencies separated by less than $310~$MHz making it possible to design a laser system able to switch between these isotopes. In our setup, this can be easily done by changing the lock-in point of each laser, as it is described in the following.

Due to the unresolved structure of the excited manyfold of the bosonic potassium isotopes, standard sub-Doppler laser cooling with light close to the $D_{2}$ line is difficult to implement and normally present a limiting temperature of more than $500~\mu$K. To overcome this limitation, schemes based on the $D_{1}$ line transitions have been successfully demonstrated with the Gray molasses procedure~\cite{Fernandes2012D140K,Salomon2013D139K,Chen2016D141K}. Therefore, \textit{cooling} and \textit{repumper} lights on the $D_{2}$ (at $767~$nm) and $D_{1}$ (at $770~$nm) lines are necessary during the experimental sequence.

In figure~\ref{fig:KLaserSystem}, we present a sketch of the potassium laser system accounting for the lights of the two $D$-lines. Three lasers from Toptica, two TA-Pro and one DLX, are used to generate the \textit{cooling} and \textit{repumper} lights from the $D_{2}$ line and the light of the $D_{1}$ line, respectively. The laser beam of each $D_{2}$ line laser is initially divided into two paths providing the lights for the 2D-MOT/\textit{imaging} and for the 3D-MOT. Double-pass AOMs are used to perform the fine frequency adjustment of the lights during the experimental sequence. The DLX laser beam is also divided in two paths in order to provide the \textit{cooling} and \textit{repumper} lights at the $D_{1}$ line.

In order to achieve the necessary laser power to efficiently cool the atoms during the MOT, we use two final stages of light amplification performed by homemade ``master oscillator power amplifiers'' (MOPAs) mounted with a \textit{tapered amplifier} (TA) chip from Eagleyard (EYP-TPA-0765-01500-3006-CMT03-0000) able to provide $1.5~$W by injecting a laser beam of only $50~$mW. In the MOPA-2D (see figure~\ref{fig:KLaserSystem}), \textit{cooling} and \textit{repumper} lights are simultaneously injected with a $50$:$50$ ratio providing $300~$mW power at the output of the optical fiber for performing the 2D-MOT and \textit{push} beams. In the MOPA-3D, not only $D_{2}$ \textit{cooling} and \textit{repumper} lights are simultaneously injected with a $10$:$1$ ratio, but also the $D_{1}$ light with a $5$:$1$ ratio, which is switched on only during the Gray molasses procedure while the $D_{2}$ light remains off. With this configuration, around $320~$mW and $210~$mW of $D_{2}$ and $D_{1}$, respectively, are obtained at the output of the 3D-MOT fiber.

The $D_{2}$ line lasers are frequency stabilized to the $D_{2}$ ground-state crossover (C.O.) of the $^{39}$K isotope by means of the saturated absorption spectroscopy technique. For selecting the potassium isotope we want to work with, AOMs are added to the saturated absorption spectroscopy scheme shifting the lock-in point of the lasers in order to cover the frequency difference between the \textit{cooling} and \textit{repumper} transitions of $308~$MHz and $108~$MHz, respectively. The DLX laser is stabilized to the $D_{1}$ ${\rm C.O.}\rightarrow \vert F'=2\rangle$ transition also of the $^{39}$K isotope. For the $D_{1}$ light, the choice between the bosonic isotopes is done combining the change in the lock-in point by $\sim 170~$MHz with the addition of a single-pass AOM to the \textit{repumper} light in order to make it resonant with the $^{41}$K transition. With this configuration, no difference in the laser powers delivered to the experiment were observed. On the experiment table, all laser beams are adjusted to have a beam diameter of $18~$mm, besides the push beam, with $d=3~$mm.

\subsection{Conservative traps to produce the two-component BEC}\label{subsec:Traps}

In this section, we discuss the trapping principles and the design of the two conservative traps we will use to produce the Bose-Bose superfluid mixture: the optically plugged Quadrupole trap~\cite{Davis1995BECNaPlug,Ketterle1996BECNaPlug} and the \textit{crossed} ODT. In figure~\ref{fig:ScienceChamberPlugODT}, we present a scheme of the configurations of both traps around the Science chamber.

\subsubsection{The optically plugged Quadrupole trap}

The optically plugged Quadrupole trap, or simply \textit{Plug} trap, is produced by a focused blue-detuned laser beam aligned through the center of a magnetic Quadrupole trap (QT), creating a barrier that repel the atoms from the region where $\overrightarrow{B}=0$, dramatically reducing the probability of Majorana losses to happen~\cite{Dubessy2012PlugKarina,Heo2010PlugNa}. The use of an ``attractive'' Plug trap, in which a red-detuned laser beam is added to the Quadrupole trap, is less favorable for sodium atoms due to the large detuning from the atomic transitions. The effective potential for the atoms in the plugged trap is:

\begin{equation}\label{eq:PlugPotential}
\begin{split}
U_{\text{Plug}} (x,y,z) = & \mu B'\sqrt{\frac{x^{2}}{4}+\frac{y^{2}}{4}+z^{2}} \\ 
& +U_{0}\frac{1}{1+(x/x_{R})^{2}}\text{e}^{-\frac{2r^{2}}{w_{0}^{2}}\frac{1}{1+(x/x_{R})^{2}}},
\end{split}
\end{equation}
where $\mu$ is the atomic magnetic moment, $B'$, the magnetic field gradient along the coils axis (in this case, chosen to be along $\widehat{z}$), $U_{0}=3c^{2}\Gamma_{0}P/\omega_{0}^{3}\Delta_{0}^{2}$ is the depth of the optical potential, $c$ is the speed of light in vacuum, $\omega_{0}$ and $\Gamma_{0}$ are, respectively, the angular frequency and the decay rate of the atomic transition, $\Delta=\omega-\omega_{0}$ is the difference between the laser and the atomic transition frequencies, named detuning of the laser, $P$ is the laser power, $w_{0}$ is the \textit{plug} beam waist at the focus, $x_{R}$ is the Rayleigh length and $r=\sqrt{(y-y_{0})^2+(z-z_{0})^{2}}$, with $(y_{0},z_{0})$ being the laser position with respect to the center of the Quadrupole trap at the $x=0$ plane.

\begin{figure}\centering
\includegraphics[width=0.45\textwidth]{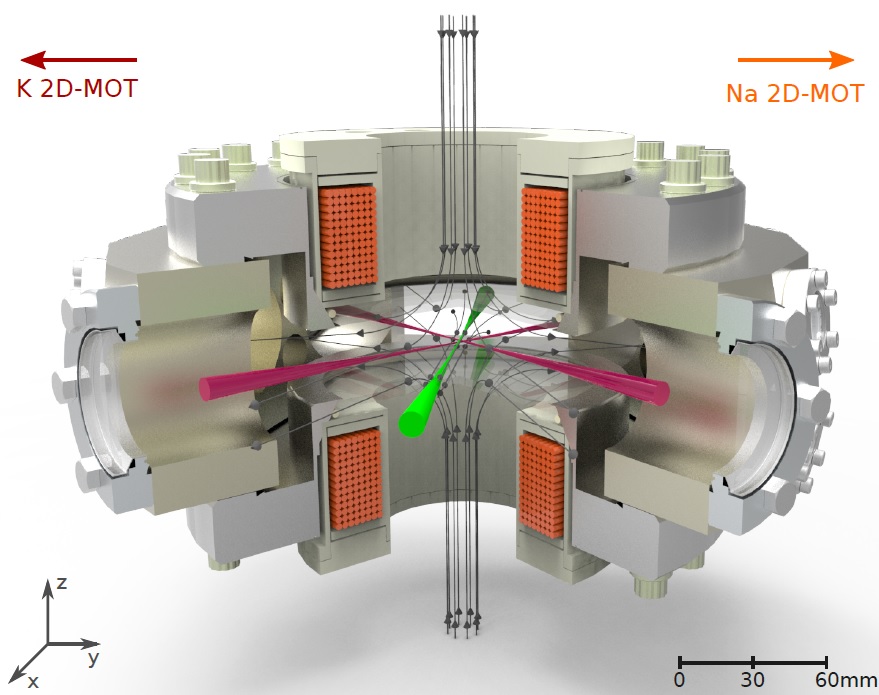}
\caption{
Vertical cut of the Science chamber with the quadrupole coils positioned along the $\widehat{z}$-axis, the \textit{plug} beam (in green) propagating along $\widehat{x}$ and aligned with the center of the quadrupole trap and the \textit{crossed} ODT laser beams (in red).
}
\label{fig:ScienceChamberPlugODT}
\end{figure}

From~\ref{eq:PlugPotential}, one can easily see that the shape of the trapping potential is strongly affected by the \textit{plug} beam alignment. Previous studies have shown that small misalignments of the \textit{plug} beam from the center of the QT (smaller than a beam waist) results in an effective potential with one single minimum~\cite{Dubessy2012PlugKarina,Heo2010PlugNa} in which the atoms would accumulate while decreasing its temperature. For small displacements around this minimum, the potential is approximately harmonic. When considering a plug beam dislocated along $\widehat{y}$, the minimum of the effective potential occurs at $(0,y_{\rm min},0)$ such that its frequencies can be given by:
%\begin{equation}
%\eqalign{\omega_{x}=\sqrt{\frac{g_{F}m_{F}\mu_{B}B'}{2my_{\rm min}}} &, ~\omega_{y}=\sqrt{\frac{4y_{\rm min}^{2}}{w_{0}^{2}}-1}\omega_{x},\cr
%& ~\omega_{z}=\sqrt{3}\omega_{x}.}
%\end{equation}
\begin{equation}
\omega_{x}=\sqrt{\frac{g_{F}m_{F}\mu_{B}B'}{2my_{\text{min}}}}, ~\omega_{y}=\sqrt{\frac{4y_{\text{min}}^{2}}{w_{0}^{2}}-1}\omega_{x}, ~\omega_{z}=\sqrt{3}\omega_{x}.
\end{equation}

\begin{figure}\centering
\includegraphics[width=0.42\textwidth]{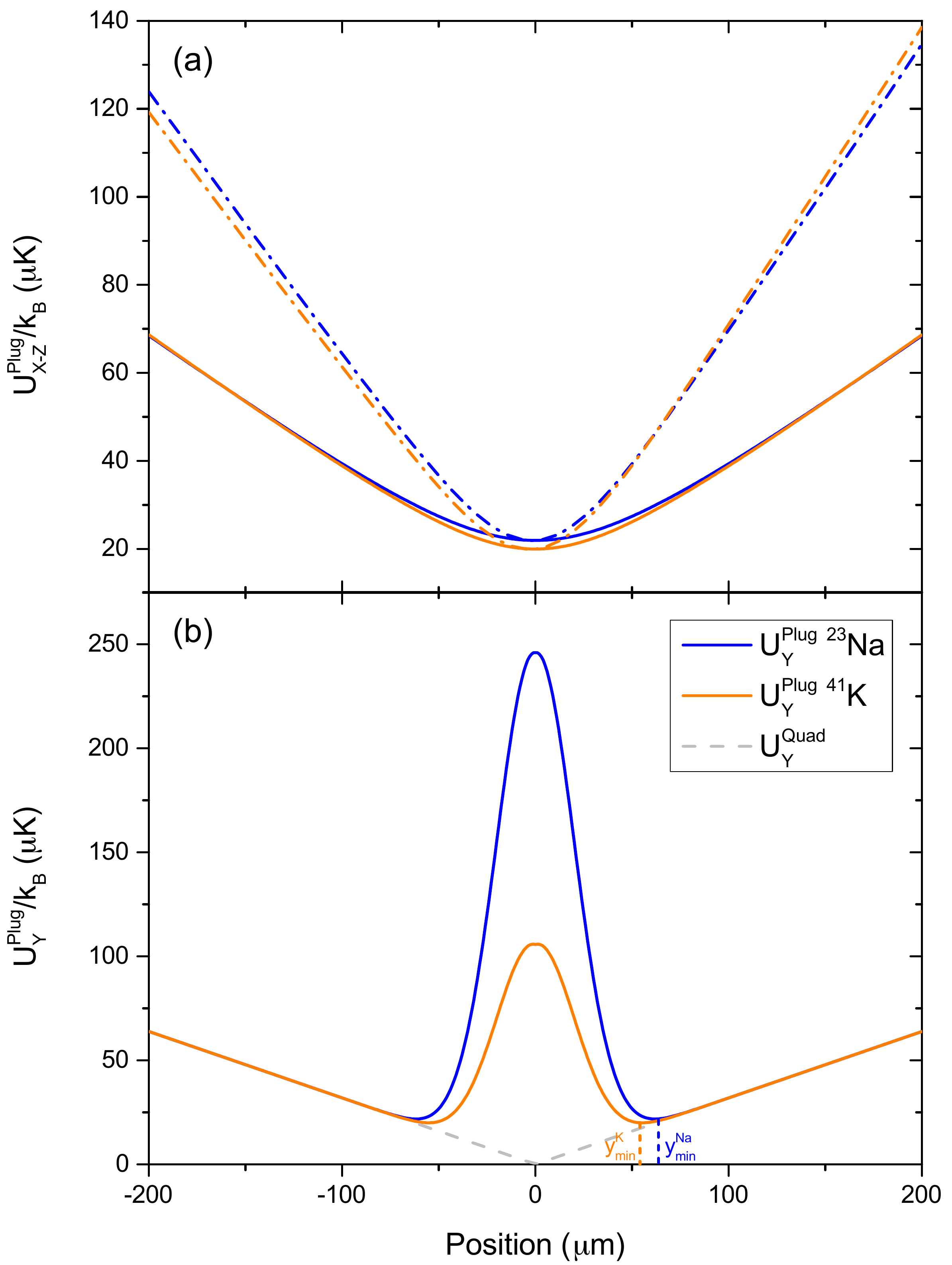}
\caption{
Plug trap potential for sodium (in blue) and for potassium atoms (in orange). The positions where the potential is minimum is different for each species as it can be seen in~(b) for the potential along $\widehat{y}$. The quadrupole potential along $\widehat{y}$ without the addition of the Plug beam is represented by the dashed gray line. The potential along the $\widehat{x}$ (solid lines) and $\widehat{z}$ (dashed lines) are displaced in~(a) and were calculated for the proper $y=y_{\rm min}$ of each atomic species.
}
\label{fig:PlugPotential}
\end{figure}

When designing a Plug trap for two different atomic species, it is important to repel both atoms from the region where Majorana losses could happen. In figure~\ref{fig:PlugPotential}~(a) and~(b), we show the designed Plug trap potential for trapping both Na and K atoms at the $\vert F=1, m_{F}=-1\rangle$ hyperfine state. For simplicity, we considered the \textit{plug} beam aligned with the center of QT in order to create the optical barrier that can be easily seen in figure~\ref{fig:PlugPotential}~(b), in which the potential along $(0,y,0)$ is plotted. For these graphs, we used a magnetic gradient of $B'=200~$G/cm, a \textit{plug} beam power of $P=4~$W and a waist of $w_{0}=43~\mu$m. The height of the barrier for the potassium atoms is a factor of $\sim 2.5$ shorter than the barrier for sodium (with $U_{0}^{\rm Na}=245~\mu$K and $U_{0}^{\rm K}=105~\mu$K), due to the larger detuning between the \textit{plug} beam and its atomic transitions. The resulting frequencies for sodium (potassium) are $f_{x}=213(172)~$Hz, $f_{y}=651(441)~$Hz and $f_{z}=368(300)~$Hz.

The magnetic field gradient is generated by a pair of circular coils in anti-Helmholtz configuration mounted along the vertical axis (see figure~\ref{fig:ScienceChamberPlugODT}) and separated by $62~$mm. These coils are also used to produce the MOT magnetic field gradient. The blue-detuned light used to produce the \textit{plug} beam is generated by a Coherent Verdi V10 laser with $\lambda=532~$nm and maximum output power of $10~$W.

\subsubsection{The crossed optical dipole trap}

Tuning the atomic interaction by means of magnetically induced Feshbach resonances~\cite{Chin2010Feshbach} consists in introducing an external \textit{bias} of magnetic field, which can only be applied in pure optical dipole traps (ODTs). In our experiment, the Bose-Bose superfluid mixture will be produced in a \textit{crossed} ODT composed of two red-detuned focused laser beams that perpendicularly cross near the minimum of the Plug trap potential (see figure~\ref{fig:ScienceChamberPlugODT}).

%\textcolor{red}{\st{Single-beam optical dipole traps are described by the second term of~}}\ref{eq:PlugPotential}. \textcolor{red}{\st{The $90^o$ \textit{crossed} ODT trapping potential is therefore obtained by summing the potentials of each individual laser beam as follows:}}
%\textcolor{red}{
%\begin{eqnarray}
%\fl U_{\rm cross} (x',y'&,z) = U_{01}\frac{1}{1+(x'/x_{R1})^{2}}{\rm e}^{-\frac{2r_{1}^{2}}{w_{01}^{2}}\frac{1}{1+(x'/x_{R1})^{2}}}\nonumber\\
%& + U_{02}\frac{1}{1+(y'/y_{R2})^{2}}{\rm e}^{-\frac{2r_{2}^{2}}{w_{02}^{2}}\frac{1}{1+(y'/y_{R2})^{2}}},
%\end{eqnarray}}
%\textcolor{red}{\st{where the indices $1$ and $2$ represent the laser beam propagating along $\widehat{x}'$ and $\widehat{y}'$, respectively. Considering small displacements from the center of the trap, the potential of the \textit{crossed} ODT is approximately harmonic }}

Single-beam optical dipole traps are described by the second term of~\ref{eq:PlugPotential}. The $90^o$ \textit{crossed} ODT trapping potential is therefore obtained by summing the potentials of each individual laser beam resulting in a approximately harmonic potential with frequencies given by $f_{i}=\sqrt{\sum_{j=1,2}f_{i,j}^{2}}$ where the $f_{i,j}$ represents the $i$-axis frequency related with the laser beam~$j$. For the particular case of identical laser beams with $U_{01}=U_{02}=U_{0}$, $w_{01}=w_{02}=w_{0}$ and $x_{R1}=y_{R2}=x_{R}$ the trapping frequencies can be simplified by:
%\begin{equation}\label{eq:FreqODT}
%\eqalign{\omega_{x',y'}= \sqrt{\frac{2U_{0}}{m}\left[\frac{1}{x_{R}^{2}}+\frac{2}{w_{0}^{2}}\right]}\approx\sqrt{\frac{4U_{0}}{mw_{0}^{2}}},~\cr
%\omega_{z}=\sqrt{\frac{8U_{0}}{mw_{0}^{2}}}},
%\end{equation}
\begin{equation}\label{eq:FreqODT}
\begin{split}
\omega_{x',y'}= & \sqrt{\frac{2U_{0}}{m}\left[\frac{1}{x_{R}^{2}}+\frac{2}{w_{0}^{2}}\right]}\approx\sqrt{\frac{4U_{0}}{mw_{0}^{2}}} \\
& \omega_{z}=\sqrt{\frac{8U_{0}}{mw_{0}^{2}}},
\end{split}
\end{equation}
which presents a ratio of $\sqrt{2}$ between the frequencies and a resulting crossed ODT depth of $U_{0}^{\rm cross}=2U_{0}$.

The \textit{crossed} ODT laser beams are generated by a MEPHISTO (MOPA-42W) laser with $\lambda=1064~$nm and maximum output power of $42~$W. The laser beam is initially divided in two parts in order to provide the two ODT laser beams desired for the \textit{crossed} configuration. The power of each beam is controlled with an AOM, which are also used as fast switchers. After the AOMs, the diffracted beam is coupled into single-mode polarization maintaining high power fibers. Around $5.5-6.0~$W exit from the fibers. The beam sizes are adjusted in order to be focused at the atoms position with $w_{0}\approx80~\mu$m. Dichroic mirrors combine the ODT beams with the in plane MOT beams.

%\begin{table}
%\caption{\label{tab:PlugParameters}Calculated parameters of the Plug trap for $^{23}$Na and $^{41}$K atoms in the $\vert F=1, m_{F}=-1\rangle$ hyperfine state. We considered the \textit{plug} beam aligned with the center of the QT and used $B'=200~$G/cm, $P=4~$W and $w_{0}=43~\mu$m.}
%\begin{indented}
%\lineup
%\item[]\begin{tabular}{@{}llll}
%\br                              
%\centering
%$\0\0\textrm{Parameter}$&$^{23}\textrm{Na}$&$^{41}\textrm{K}$\cr 
%\mr
%\0\0$y_{\textrm{min}}$&$64~\mu\textrm{m}$&$55~\mu\textrm{m}$\cr
%\\
%\0\0$f_{x}$&$213~\textrm{Hz}$&$172~\textrm{Hz}$\cr
%\\
%\0\0$f_{y}$&$651~\textrm{Hz}$&$441~\textrm{Hz}$\cr
%\\
%\0\0$f_{z}$&$368~\textrm{Hz}$&$300~\textrm{Hz}$\cr
%\\
%\0\0$U_{0}$&$245~\mu\textrm{K}$&$105~\mu\textrm{K}$\cr
%\\
%\0\0$U_\textrm{Plug}(0,y_{\textrm{min}},0)$&$22~\mu\textrm{K}$&$20~\mu\textrm{K}$\cr
%\\
%\end{tabular}
%\end{indented}
%\end{table}

\section{The two-species MOT}\label{sec:TwoMOTs}

In this section, we describe the individual magneto-optical traps for sodium and potassium atoms and characterize the operation of the combined two-species MOT.

\subsection{Sodium 2D and 3D-MOTs}\label{subsec:Na2De3DMOTs}

The 3D-MOT of sodium in the SC is loaded from the pre-cooled atoms coming from the \textit{modified} 2D-MOT (2D-MOT + adapted Zeeman slower) performed in the Na 2D-MOT chamber.

Around $200~$mW ($I \sim 20~I_{\textrm{s}}$, with $I_{\textrm{s}}$ being the saturation intensity $I_{s}=6.26~$mW/cm$^2$) exits the 2D-MOT fiber with a central frequency near the cooling transition ($\delta_{\textrm{cool-2D}}\sim-3.3~\Gamma_{\textrm{Na}}$, where $\Gamma_{\textrm{Na}}=9.79~$MHz is the linewidth of the D$_{2}$ line transitions) and sidebands at $1.712~$GHz, providing $\delta_{\textrm{rep-2D}}\sim-3.4~\Gamma_{\textrm{Na}}$. This light is divided into two circularly polarized retro-reflected laser beams which perpendicularly cross in the center of the Na 2D-MOT vacuum chamber. The two-dimensional magnetic quadrupole field used for the 2D-MOT is produced by four sets of nine permanent magnets each (K$\&$J Magnetic Inc. model BX082 with dimensions of $(3.2,12.7,25.4)~$mm and magnetization $\textbf{M}=(10.5\times 10^{5},0,0)$) with separations $d_{y}=6~$cm and $d_{z}=9~cm$, resulting in a magnetic field gradient along the beam's propagation directions of $60~$G/cm. Along the vertical direction, in which travels the atomic flux coming from the oven, the residual magnetic field, together with a laser beam counter-propagating with the atomic flux, is used to perform an adapted Zeeman slower~\cite{Tiecke20092DMOTLi,Lamporesi20132DMOTNa,Pedrozo2016Comparison}. Around $150~$mW ($\sim 15~I_{\textrm{s}}$) is used for the Zeeman laser beam with cooling and repumper lights red-detuned from the atomic transitions by $\sim 26~\Gamma_{\textrm{Na}}$. Its polarization is set to be along the long axis of the 2D-MOT which, in combination with the magnetic field, results in a circularly polarized light for the atoms traveling along $\widehat{z}$~\cite{Lamporesi20132DMOTNa}. Finally, the pre-cooled atoms in the 2D-MOT are guided to the SC by the use of a \textit{push} beam with waist $w_{0}\sim 500~\mu$m, power of $300~\mu$W ($\sim 0.03~I_{\textrm{s}}$) and frequency blue-detuned from the cooling transition ($\delta_{\textrm{push}}\sim + 1.2~\Gamma_{\textrm{Na}}$).

The 3D-MOT of sodium in the SC operates in the Dark-SPOT MOT configuration~\cite{Ketterle1993DarkSPOT}. We use a linear gradient of magnetic field ($\sim 11~$G/cm) produced by the Quadrupole coils, three circularly polarized retro-reflected laser beams with frequency near the cooling transition ($\delta_{\textrm{cool-3D}}\sim-1.7~\Gamma_{\textrm{Na}}$) and one single pass laser beam near the repumper transition ($\delta_{\textrm{rep}}\sim-0.8~\Gamma_{\textrm{Na}}$). The repumper beam has a ``hole'' (dark-SPOT) at the center of its intensity profile which is imaged into the atoms by a $1$:$1$ telescope, such that only cooling light arrives in the center of the MOT. Therefore, the atoms in the central region accumulate in the $F=1$ ground-state and stop to interact with the cooling light, decreasing the re-scattered radiation and increasing the density of the atomic cloud. In our experiment, the dark-SPOT is produced by a circular ``mirror-mask'' with $5.4~$mm of diameter, chosen in order to maximize the phase-space density ($\propto N/T^{3/2}$) of the atoms transferred to the magnetic trap. After a loading time of $5~$s, around $5\times 10^{9}~$atoms at $350~\mu$K are trapped in the dark 3D-MOT.

\subsection{Potassium 2D and 3D-MOTs}\label{subsec:K2De3DMOTs}

The 3D-MOT of potassium atoms in the SC is loaded from the pre-cooled atoms coming from the 2D-MOT performed in the K 2D-MOT chamber operating as a vapour cell. Thanks to the larger natural abundance of the $^{39}$K isotope, we choose this isotope to perform the initial characterization of the system. Besides the atom number, we do not expect considerable changes in switching to the $^{41}$K isotope~\cite{39K-41K}.

Around $300~$mW ($\sim 135~I_{\textrm{s}}$ with $I_{s}=1.75~$mW/cm$^{2}$) containing lights near the cooling ($\delta_{\textrm{cool-2D}}\sim-3.5~\Gamma_{\textrm{K}}$, where $\Gamma_{\textrm{K}} = 6.03~$MHz is the linewidth of the D$_{2}$ line transitions) and the repumper ($\delta_{\textrm{rep-2D}}\sim-3.0~\Gamma_{\textrm{K}}$) transitions in a $50$:$50$ ratio exits the 2D-MOT fiber. A small part of this light ($\sim 0.3~$mW) is used for the \textit{push} beam while its main part is divided into two circularly polarized retro-reflected laser beams which perpendicularly cross in the center of the K 2D-MOT chamber. The two-dimensional magnetic quadrupole field is produced by four sets of four permanent magnets each (the same model used for the Na 2D-MOT) producing a magnetic field gradient along the beam's propagation direction of $30~$G/cm. The pre-cooled atoms are transferred to the SC guided by the \textit{push} beam. The resulting atomic flux is of around $2-4\times 10^{7}~$atoms/s.

The 3D-MOT of $^{39}$K is produced with a linear gradient of magnetic field of $16~$G/cm and a total laser power of $320~$mW ($\sim 144~I_{\textrm{s}}$) containing lights near the cooling ($\delta_{\textrm{cool-3D}}\sim-5~\Gamma_{\textrm{K}}$) and the repumper ($\delta_{\textrm{rep-3D}}\sim -6.5~\Gamma_{\textrm{K}}$) transitions in a $10$:$1$ ratio. The potassium MOT also operates in a retro-reflected configuration. After a loading time of $8~$s, around $2\times 10^{8}~$atoms at $5.5~$mK are trapped in the MOT.

\begin{figure}\centering
\includegraphics[width=0.48\textwidth]{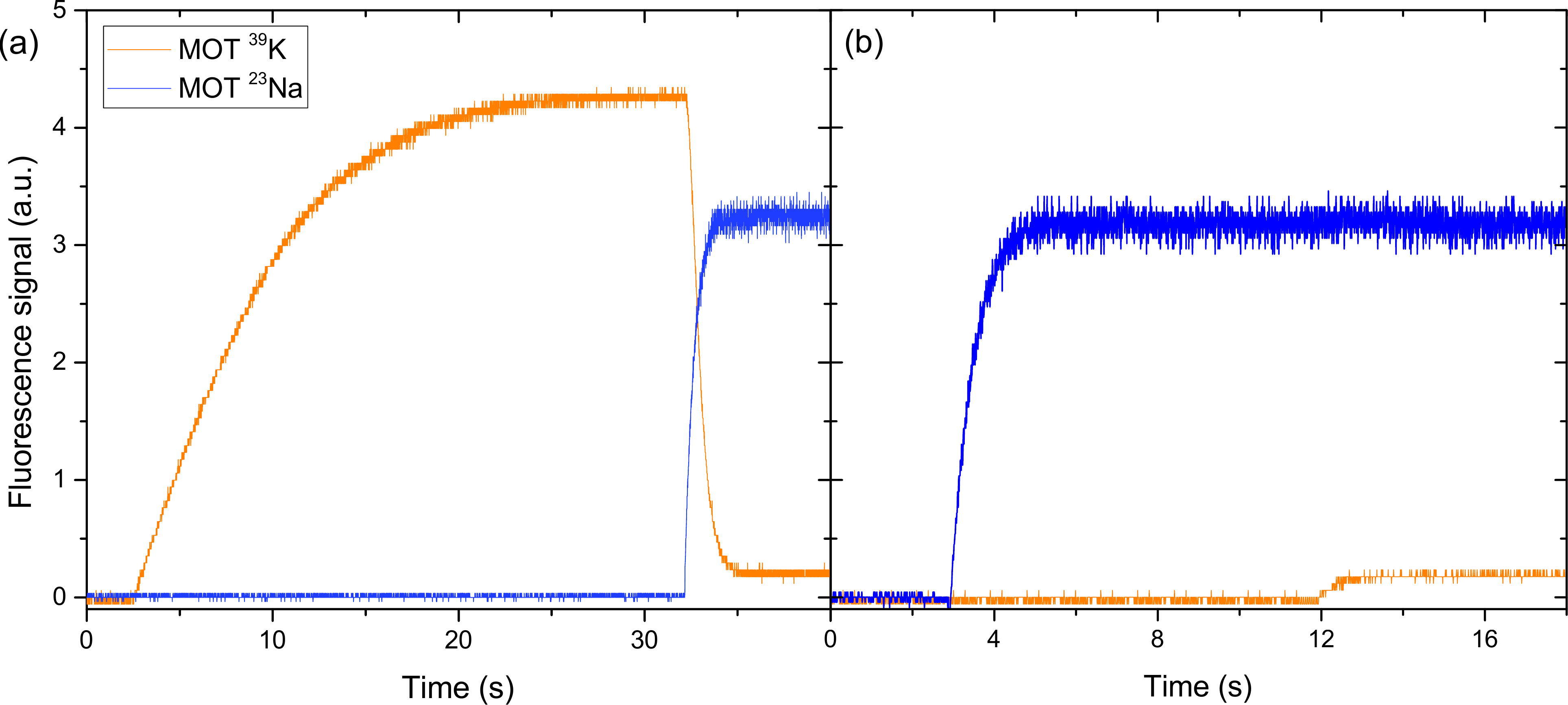}
\caption{Fluorescence signals from the potassium ($^{39}$K, in orange) and sodium (in blue) MOTs as a function of the loading time. In (a), we show the dramatic decrease on the fluorescence signal of the potassium MOT as soon as the MOT of sodium is loaded. This decrease represents an atom loss of more than one order of magnitude (from $\sim 2\times 10^{8}~$atoms to $\sim 10^{7}~$atoms, measured with absorption imaging). In (b), we show the opposite situation in order to observe that the fluorescence signal from the sodium MOT does not change in the presence of the potassium atoms. 
}
\label{fig:TwoSpeciesMOT}
\end{figure}

\subsection{The $^{23}$Na-$^{39}$K MOT}\label{subsec:TwoMOTs}

The first step in a two-species ultracold atom experiment consists in creating an efficient method to load and cool the different species prior the transference to a conservative trap. The easiest approach is to perform a two-species MOT simultaneous loading both species. However, light induced losses~\cite{Weiner199ReviewLossesMOT} could be a limiting factor while operating the two-species MOT.

In the case of the $^{23}$Na-$^{39}$K MOT, we have observed this kind of losses. Due to the large atom number difference between the individual MOTs, the most affected species is the potassium, contrary to the previous observations in which the lightest species suffers the stronger losses. In figure~\ref{fig:TwoSpeciesMOT}, we show the fluorescence signal of the sodium and potassium MOTs captured by independent photodiodes (sensitive to light at $767~$nm or at $589~$nm) for two different loading configurations: (a) the potassium MOT is fully loaded before turning on the sodium MOT, showing the dramatic decrease in the potassium fluorescence; and (b) with the opposite situation, in which no difference in the sodium fluorescence was observed. The drastic drop in the K MOT fluorescence corresponds to a decrease of more than one order of magnitude in the number of atoms (from $\sim 2\times 10^{8}~$atoms to $\sim 10^{7}~$atoms) making it difficult to perform the subsequent cooling procedures of the potassium cloud.

In order to circumvent the atom loss resulted from the light induced losses, one can implement a ``Two-stage MOT loading'' approach in a similar way as the one described by C.-H. Wu in~\cite{Wu2013Thesis}. It consists in performing the MOT of each species at different times of the experimental sequence. Since this method has been demonstrated to be efficient for loading $^{23}$Na and $^{40}$K atoms, we are confident that it will work also in the case of switching for the potassium bosonic isotopes, $^{39}$K and $^{41}$K. The first attempts on this direction are under development in the laboratory. Moreover, the ``Two-stage MOT loading'' could be improved with a Gray molasses cooling of the sodium atoms~\cite{Colzi2016GrayMolNa} producing a colder cloud before loading the potassium MOT.

\section{Production of a $^{23}$Na BEC}\label{sec:NaBEC}

In this section, we describe the experimental sequence for producing a BEC of sodium with $1\times 10^{6}~$atoms.

\subsection{From the 3D-MOT to the magnetic trapping}\label{subsec:MOTtoPlugTrap}

%In order to achieve the high phase-space densities required approaching degenerate systems, the atoms must be transferred to a conservative trap in which the evaporative cooling procedure~\cite{Ketterle1996BECNaPlug} can finally take place. The transference of the atoms from the MOT to such conservative traps is one of the most critical steps in BEC experiments and the addition of intermediate stages can be used to improve its efficiency.

We perform two intermediate steps before transfering the atoms from the $^{23}$Na dark 3D-MOT to the Plug trap: a \textit{dark}-molasses and a pre-pump stage. The first stage, used to cool the atoms below the Doppler limit, is done combining the usual bright molasses~\cite{Phillips1998NobelLecture} with the Dark-SPOT technique. At the beginning of the \textit{dark}-molasses, the magnetic field is abruptly turned off and the detuning and power of the cooling light are ramped during $1.5~$ms up to their final molasses values ($\delta_{\textrm{cool-Mol}}\sim-2.6~\Gamma_{\textrm{Na}}$ and $I_{\textrm{cool-Mol}}\sim 2.5~I_\textrm{s}$), which are kept constant during aditional $3~$ms. The repumper light (detuning and power) is not changed during this stage. In the end of the \textit{dark}-molasses we obtain $5\times 10^{9}~$atoms at $80~\mu$K.

After the \textit{dark}-molasses, we perform the pre-pump stage by turning off the repumper light used for the dark-SPOT and letting the atoms expand during $250~\mu$s in the presence of only cooling light. Since the cooling transition is not closed, after a few cycles all the atoms are in the $\vert F=1\rangle$ manifold equally distributed between the three Zeeman sublevels.

\subsection{Evaporative cooling in the Plug trap}\label{subsec:NaPlugTrap}

The transference of the pre-pumped $^{23}$Na atoms to the Quadrupole trap is done by abruptly switching on the magnetic field at $B'_{\textrm{catch}}=94~$G/cm. This value was optimized as the lowest magnetic field gradient that allows all trappable atoms to remain trapped into the magnetic trap. After that, the field is adiabatically ramped up to its maximum value $B'_{QT}=190~$G/cm during $500~$ms. Once the atoms are trapped in the pure Quadrupole trap, the \textit{plug} beam is turned on by ramping the laser light intensity from zero to $P_{\textrm{Plug}}=2.10~$W in $100~$ms~\cite{PlugComments}. Around $1.5\times 10^{9}~$atoms at $220~\mu$K are transferred to the Plug trap with a lifetime of $46~$s.

\begin{figure}\centering
\includegraphics[width=0.48\textwidth]{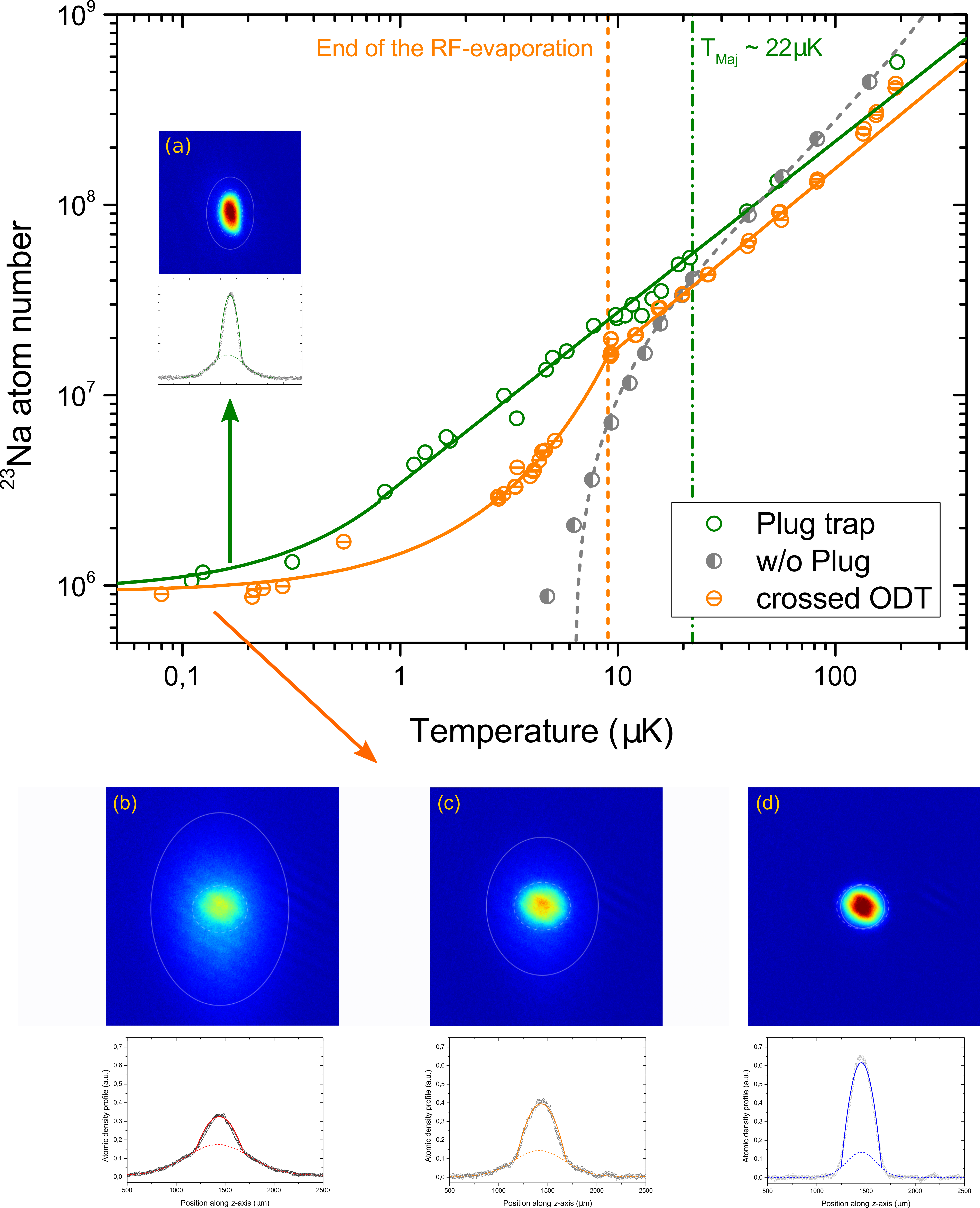}
\caption{
Log-log graph of $N~vs.~T$ for different points of the RF-evaporation of the sodium atoms. The optimized evaporation procedure in the Plug trap (green circles), the same frequency ramp performed only in the Quadrupole trap (gray circles) and the complete evaporation sequence for the BEC in the crossed ODT (orange circles) are displayed in the graph as well as the Majorana temperature (green dashed line with $T_{\textrm{Maj}}\approx 22~\mu$K) identified as the temperature for which the two data points starts to deviate and the point in which the atoms are transferred to the ODT (orange dashed line).
}
\label{fig:NxTPlug}
\end{figure}

With the atoms in the Plug trap, forced radio-frequency (RF) evaporation is applied selectively removing the hottest atoms~\cite{Mauhara1988RFEvaporation}. The RF radiation is delivered by a two-loop circular antenna with diameter of $38~$mm placed below the upper Quadrupole coil. The evaporation ramp for producing a BEC of $^{23}$Na in the Plug trap lasts $12.3~$s and it is made of a series of linear ramps resembling an exponential ramp starting with $f_{\textrm{RF}}=45~$MHz and final frequency between $1.0-0.3~$MHz. The efficiency of the evaporation procedure can be evaluated by plotting $N~vs.~T$ in a log-log scale as it is shown in the graph of figure~\ref{fig:NxTPlug}. Two different set of data, with (green open circles) and without (gray half-filled circles) the addition of the \textit{plug} beam, are displayed in the graph while the solid curves are just used as guides for the eyes. The Majorana temperature is identified as the temperature for which the two set of data points start to deviate due to the occurrence of Majorana losses and it is around $22~\mu$K for the sodium atoms in our system. For the green circles, the efficiency of the evaporation procedure can be estimated by the parameter $s=\log(N)/\log(T)$ with $s\leq1$ for achieving run-away evaporation. During almost all the RF-evaporation sequence $s=0.89$, ensuring its efficiency. The onset of BEC is identified by the appearance of a bimodal density distribution of the atomic cloud observed after $20~$ms of time-of-flight (see the inset~(a) of figure~\ref{fig:NxTPlug}) for temperatures below $T_{c}=1.2~\mu$K and $f_{\textrm{RF}}=1.0~$MHz. Almost pure BECs with $8\times 10^{5}~$atoms could be easily produced in the Plug trap by reducing even further the RF-frequency between $0.5-0.4~$MHz.

\subsection{Final cooling in the crossed ODT}\label{subsec:NaBECODT}

The transfer to the \textit{crossed} ODT is done after performing the RF-evaporation procedure, as described in the previous section, until $f_{\textrm{RF}}=2.0~$MHz and $t_{\textrm{Evap}}=9.4~$s. The two \textit{crossed} ODT laser beams are adiabatically switched on to its maximum power ($\sim 6~$W) during the final $400~$ms of the RF-evaporation. Later, the magnetic field is ramped down during $500~$ms and the \textit{plug} beam is abruptly switching off after the end of the magnetic field ramp. Around $6 \times 10^{6}~$atoms at $\sim 6~\mu$K are trapped in the \textit{crossed} ODT.

With the atoms in the \textit{crossed} ODT, an optical evaporation can continue to cool the atomic sample by slowly decreasing the ODT depth. In our experiment, this done by reducing the ODT laser beams powers with a series of six linear ramps for each beam resembling an exponential decay. The total optical evaporation procedure lasts $3.9~$s during which the power is finally reduced by a factor of $1/10$ for the ODT1 ($P_{1}^{\textrm{final}}\approx 610~$mW) and of $1/20$ for the ODT2 ($P_{2}^{\textrm{final}}\approx 290$mW) with respect to each initial power. The final ODT depth for $^{23}$Na atoms is $U_{0}^{\textrm{cross}}\approx 3.54~\mu$K and the resulting frequencies are $f_{x}=80(2)~$Hz, $f_{y}=106(2)~$Hz and $f_{z}=128(1)~$Hz, measured combining the parametric heating technique~\cite{Dubessy2012PlugKarina} with the excitation of dipolar oscillations of the atomic cloud. All frequencies show a good agreement with the frequencies obtained with~\ref{eq:FreqODT}. The route of the complete evaporation procedure for achieving the BEC in the crossed ODT (forced-RF evaporation and optical evaporation) can be seen on the log-log graph of $N~vs.~T$ presented in figure~\ref{fig:NxTPlug}.

%\begin{figure}\centering
%\includegraphics[width=0.38\textwidth]{Figures/figure9a.pdf}
%\includegraphics[width=0.38\textwidth]{Figures/figure9b.pdf}
%\caption{
%Crossed ODT trapping frequencies measured combining parametric~(a) and dipolar~(b) oscillations of the atomic cloud. The dipolar oscillations~(a) are excited by abruptly switching off the laser beams of the crossed ODT letting the atoms fall under gravity during $1.5~$ms before re-capturing them in the ODT. The parametric oscillations~(b) are excited by modulating the laser power of the ODT laser beams with a ratio similar to the one used for the current of the Quadrupole field, in Sec.~\ref{subsec:NaPlugTrap}. The frequency along gravity with $f_{z}=128~$Hz can be measure using both approaches, while the in-plane frequencies, $f_{x}=80~$Hz and $f_{y}=106~$Hz, are only measured with the parametric heating.
%}
%\label{fig:ODTFReq}
%\end{figure}

The achievement of the $^{23}$Na BEC in the \textit{crossed} ODT is also revealed with the appearance of a bimodal density profile observed after $20~$ms of time-of-flight as illustrated in figure~\ref{fig:NxTPlug}~(b-d) for different steps of the optical evaporation. Almost pure BECs with $\sim 1\times 10^{6}~$atoms are obtained by the end of the optical evaporation with lifetimes as long as $14~$s.

\section{Production of a vortex lattice in the BEC of sodium}

In the direction of producing coupled vortex lattices and binary quantum turbulence in the Bose-Bose superfluid mixtures produced in our setup, we developed a large numerical aperture imaging system combined with a versatile stirring beam setup enabling the design of a large variety of stirring patterns. In this section, we describe the technical details of these two setups and profs its efficiency by producing a vortex lattice in the BEC of sodium.

\subsection{Vertical imaging and stirring setup}\label{subsec:ImgSetup}

\begin{figure}\centering
\includegraphics[width=0.45\textwidth]{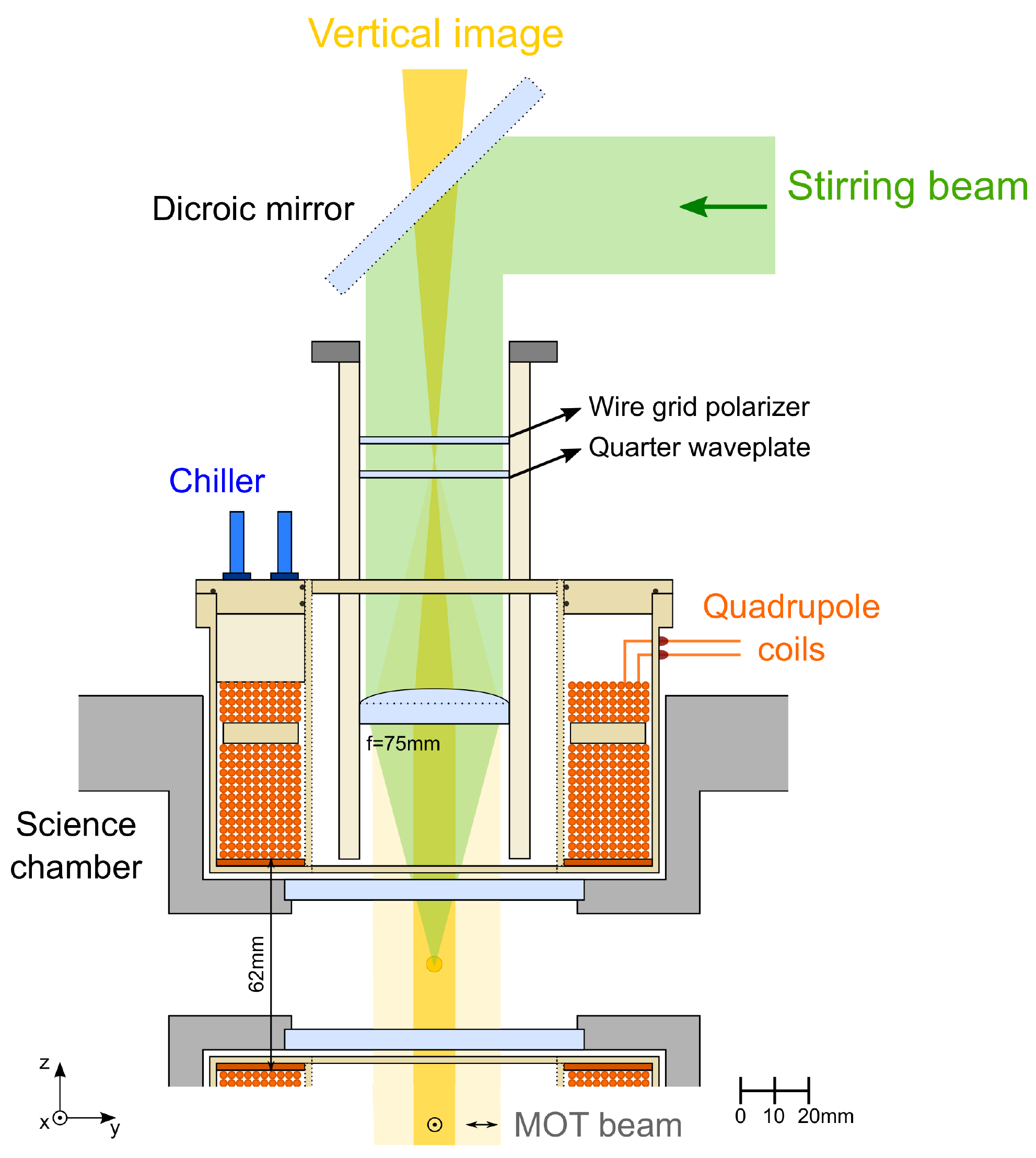}
\caption{Cross-section view of the vertical MOT beams (light yellow), the imaging beam (dark yellow) and the stirring beam (green) around the science chamber (gray). The position of the BEC is indicated by a yellow circle in the center of the SC. The polarization of the MOT, imaging and stirring beams are adjusted such that the wire-grid polarizer reflects the MOT beam and transmits the vertical imaging and the stirring beams.
}
\label{fig:VerticalImagingSetup}
\end{figure}

The stirring technique~\cite{Madison2000RotatingBEC,AboShaeer2001VortexLattice,Raman2001StirringBEC} vastly used in ultracold atom experiments consists in the addition of a tightly focused blue-detuned laser beam whose rotates along the symetry axis of the BEC. The nucleated vortices are oriented with the rotation axis and an imaging beam co-propagating with the stirring beam is normally used to observe the vortex lattice. In our experiment, the symmetry axis of the $^{23}$Na BEC produced in the crossed ODT is along the gravity direction and we use a vertical setup combined with the MOT beams for imaging and stirring. A scheme of the vertical setup along the science chamber is illustrated in figure~\ref{fig:VerticalImagingSetup}. A thin wire-grid polarizer reflects the vertical MOT beam and transmits the stirring and the imaging beams without significant deformation.

The large clear aperture of the viewports along the $\widehat{z}$-axis enables the design of a high resolution setup. A two-inch achromatic 75~mm focal length lens (Thorlabs AC508-075-A) is used as a simple microscope objective. Despite its simplicity, we have measured an optical resolution $R \approx 1~\mu$m for $532~$nm and $589~$nm, stirring and imaging wavelengths, respectively. This optical resolution gives the possibility of producing highly localized potentials with the green light. In the future, a custom high numerical achromatic microscope will be installed covering the range of the different wavelengths necessary for detecting the Bose-Bose mixture. A custom designed dichroic mirror from Laseroptik combines the vertical imaging and the stirring beams. The vertical imaging magnification equal to $\times 4$ is obtained by adding a $300~$mm focal length lens before the CCD, ensuring a good visualization of the vortices during time-of-flight.

\begin{figure}\centering
\includegraphics[width=0.48\textwidth]{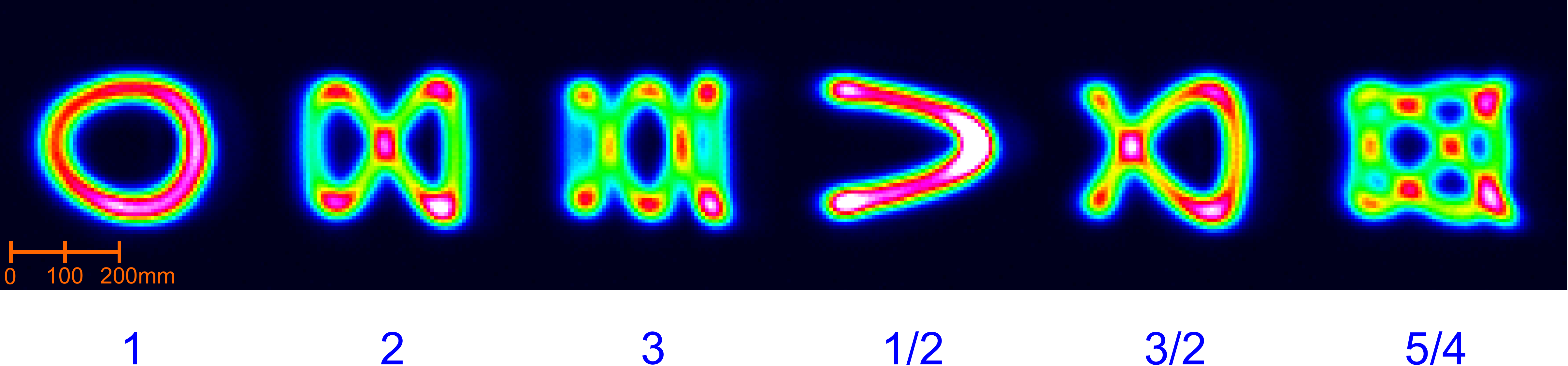}
\caption{Lissajous figures produced by the stirring beam and imaged into a CCD camera with long exposure time. The ratio between the modulation frequencies of each AOM was varied and it is indicated by the number below each image.
}
\label{fig:Lissajous}
\end{figure}

The stirring beam produced with light at $532~$nm is focused to a beam waist of $7~\mu$m in the atomic plane. In order to stir the laser beam, two crossed AOMs are used to produce a two-times diffracted beam, allowing a in-plane control of the stirring beam position. By modulating the RF-frequency that control the AOMs with frequency $f_{\textrm{Stirr}}$, our system is able to produce the different Lissajous figures shown in figure~\ref{fig:Lissajous}. The circle on the left of the image is the usual pattern used for stirring and it is the one we applied to the atoms in the results presented in the next section.

\subsection{Nucleation of a vortex lattice}\label{subsec:VortexLattice}

\begin{figure}\centering
\includegraphics[width=0.45\textwidth]{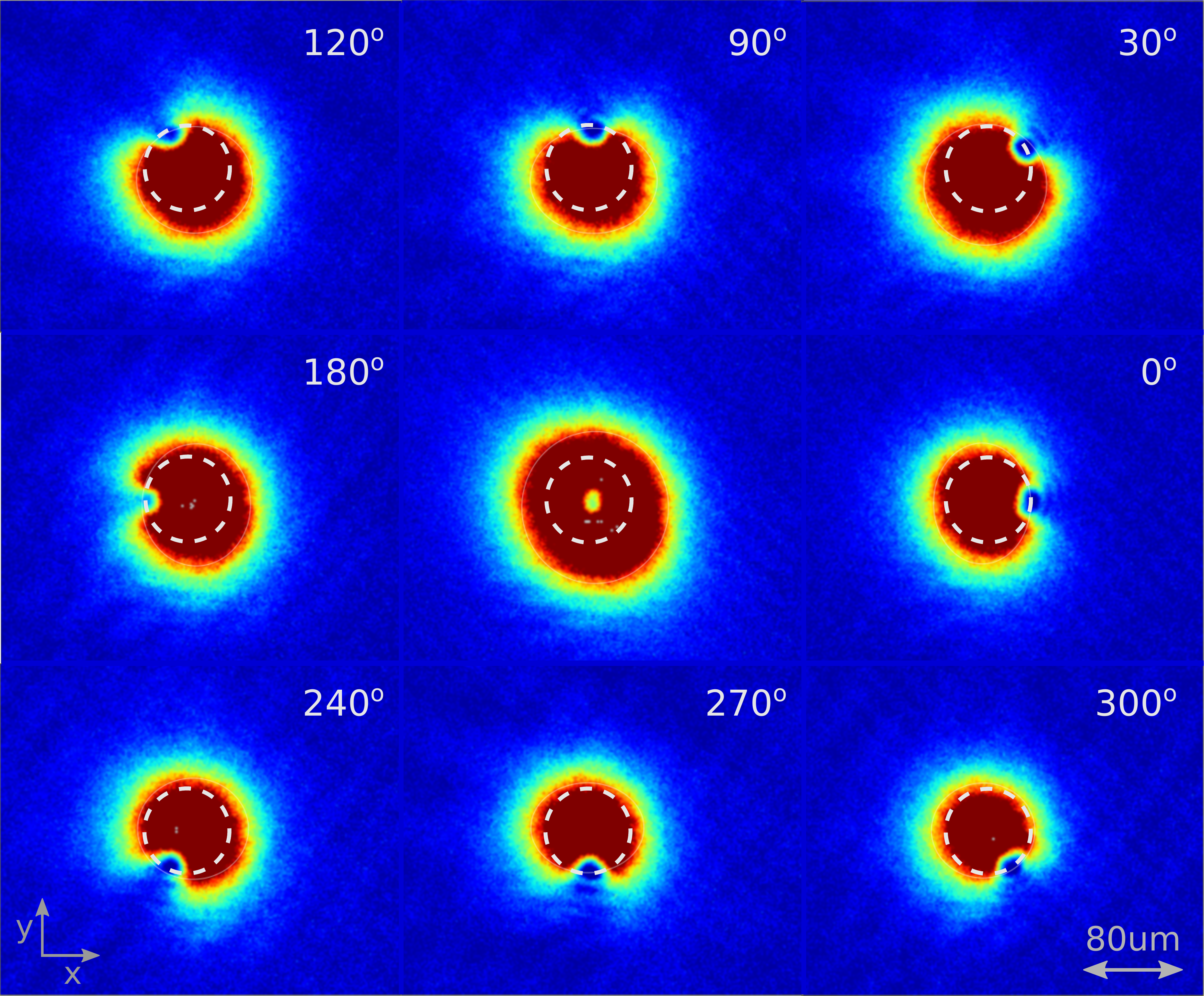}
\caption{Mapping of the stirring beam position in the atomic cloud while simulating a circle with radius $\sim 40~\mu$m. The central image show the stirring positioned at the center of the cloud.
}
\label{fig:MapStirring}
\end{figure}

The position of the stirring beam in the atomic plane can be directly measured by doing an \textit{in situ} image of the atoms in the crossed ODT in the presence of the stirring. The repulsive potential that it creates results in a density dip in the atomic density profile. Figure~\ref{fig:MapStirring} shows a series of absorption images taken while simulating a circular path with radius of $\sim 40~\mu$m, used to nucleate vortices.

\begin{figure}\centering
\includegraphics[width=0.4\textwidth]{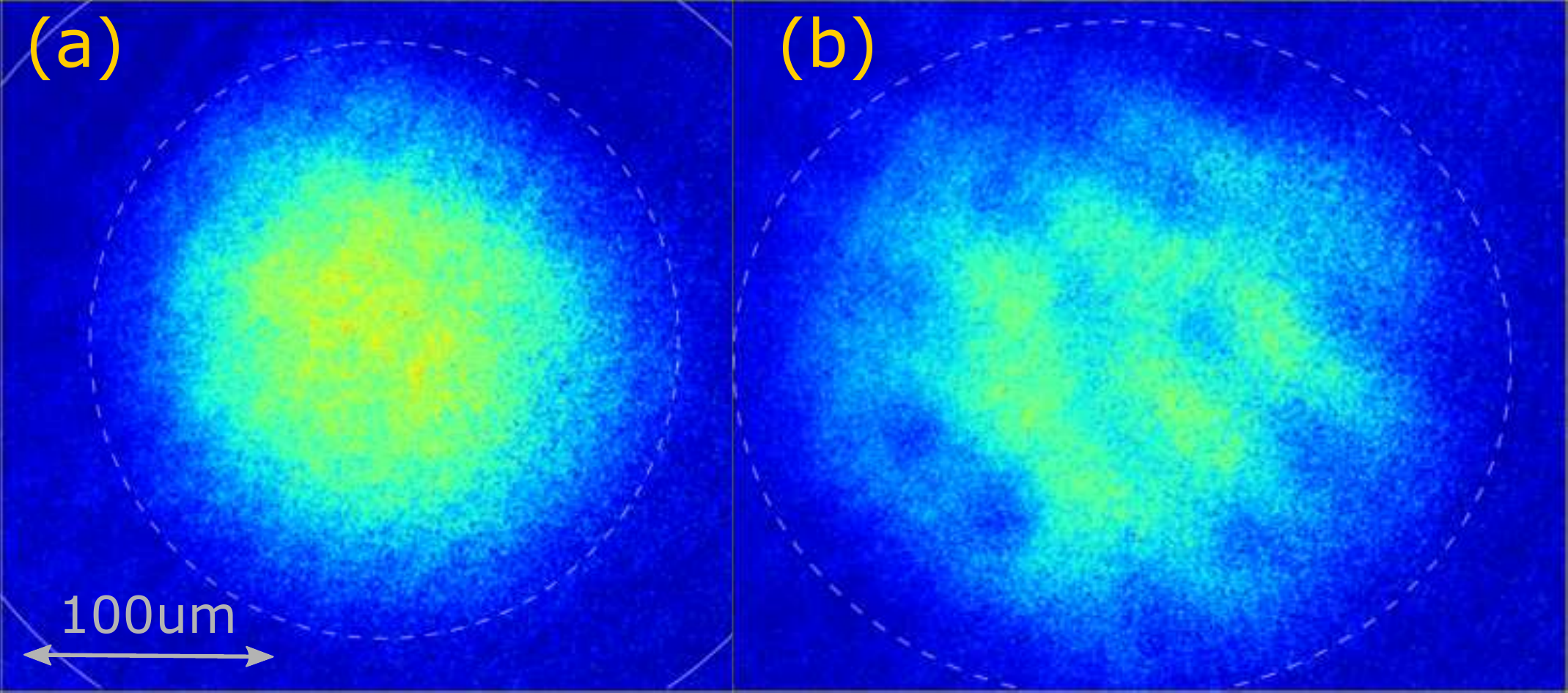}
\caption{Comparison between the usual BEC of sodium obtained in the crossed dipole trap~(a) and the stirred BEC~(b) with a vortex lattice.
}
\label{fig:VortexLattice}
\end{figure}

The nucleation of a vortex lattice with the stirring beam is obtained with a circular path with radius of $8~\mu$m and different modulation frequency values, ranging from 40 to 85~Hz. After achieving an almost pure BEC of $^{23}$Na atoms in the end of the optical evaporation, the stirring beam power is adiabatically ramped up from 0 to $75~\mu$W in 100~ms. During this time, the stirring beam is already rotating around the BEC. The BEC is rotate for another 1~s at full stirring power, before decreasing it during other 10~ms. The atomic cloud remains trapped in the ODT for another $100~$ms before being released and imaged after 30~ms of time-of-flight. After this time, the size of the vortices have sufficiently expanded in order to a clear vortex lattice to be visible with our imaging system. In figure~\ref{fig:VortexLattice}, we show the resulting vortex lattice for $f_{\textrm{Stirr}} = 80~$Hz~(b) in comparison with the static $^{23}$Na BEC~(a).

The number of nucleated vortices increases approximately linearly with the rotation frequency approaching the limiting case of rigid body rotation~\cite{Raman2001StirringBEC}. On the other hand, the centrifugal force decreases the in-plane confinement creating an upper bound for $f_{\textrm{Stirr}} = f_{r}$ for which the atoms are no longer confined. In our experiment, frequencies up to $f_{\textrm{Stirr}}=85~$Hz do not show any atom loss and by changing the stirring time, vortex lattices with 14~vortices were produced with a lifetime of $\sim 400~$ms. Further investigations on the decay of the vortex lattice are on going to rule out the contributuion of the residual thermal component.

\section{Conclusions and Perspectives}

In this paper, we have described a new experimental machine for studying Bose-Bose superfluid mixtures composed of $^{23}$Na and the bosonic isotopes of potassium atoms ($^{39}$K and $^{41}$K). The versatility of the experimental apparatus surely opens up new prospects to the study of Bose-Bose mixtures.

The first step in simultaneously trapping and cooling the two atomic species was performed with a two-species MOT. Light induced losses strongly affected the potassium atoms preventing a successful transference of the atoms to the Plug trap. Next, we described the experimental sequence for producing an almost pure $^{23}$Na Bose-Einstein condensate in a \textit{crossed} optical dipole trap with $1\times 10^{6}~$atoms and lifetimes as long as $14~$s. We described the implementation of a high resolution setup for imaging and stirring. A vortex lattices with up to 14 vortices and lifetimes of $\sim 400~$ms were produced attesting the efficiency of the experimental apparatus in studying the dynamics of vortices.

Further experiments focused on the role played by the residual thermal component in the nucleation and decay of the vortex lattice as well as the search for new methods of producing quantum turbulence~\cite{Allen2014QT8stirring} are of great importance. In the case of the Na-K bosonic mixtures, the presence of heteronuclear Feshbach resonances~\cite{Viel2016NaKFeshPredictions} will allow to explore different new scenarios~\cite{Castilho2017Thesis}. In particular, the dynamics of coupled vortex lattices could be studied over the different miscibility regimes allowing the observation of exotic vortex configurations~\cite{Kasamatsu2003RotTwoBECs,Mason2011RotTwoBECs,Kuopanportti2012ExoticVortexLattices}. Relying on the large difference between the stirring potential for sodium and potassium, a species selective vortex nucleation method could also be realized in order to study the transfer of vorticity between two superfluids.
\\

\section{Acknowledgements}
The authors thank R. F. Shiozaki for developing the experimental control program and V. Monteiro and D. V. Magalh\~{a}es for providing experimental support. This work was supported by the financial agencies CNPQ, CAPES and by the grant $\# 2014/14198-0$ from S\~{a}o Paulo Research Foundadion (FAPESP).


\begin{thebibliography}{}
\bibitem{39K-41K} Even the number of atoms can be kept the same by increasing the background pressure of the K 2D-MOT chamber from $1\times 10^{-9}~$Torr to $1-2\times 10^{-8}~$Torr, closer to the usual values used for performing 2D-MOTs of potassium atoms.
\end{thebibliography}
\end{document}